\documentclass[12pt]{article}
\usepackage{amsmath}

\def\hybrid{\topmargin -20pt    \oddsidemargin 0pt
        \headheight .4in \headsep 0pt
        \textwidth 6.25in       % A4 paper
        \textheight 9.1in       % A4 paper
        \marginparwidth .875in
        \parskip 5pt plus 1pt   \jot = 1.5ex}
\hybrid

\numberwithin{equation}{section}
\numberwithin{table}{section}

\newcommand{\ap}{{\alpha'}}

\begin{document}

\begin{titlepage}

  \begin{center}
    {\Large\textbf{A Note on Moduli Stabilisation in Heterotic
        \\[.2cm] Models in the Presence of Matter Fields}}

    \vskip1cm

    \textbf{Andrei Micu}
    \vskip.5cm
    \textit{Department of Theoretical Physics \\ National Institute
      for Physics and Nuclear Engineering -- Horia Hulubei \\
      Str. Atomi\c{s}tilor 407, P.O.~Box MG-6 M\u{a}gurele, Romania}\\
    \texttt{amicu@th.physik.uni-bonn.de} 

    \vspace{1cm}
    \textbf{Abstract}
  \end{center}

  \noindent In this brief note we analyse a toy model which can
  be derived from heterotic string compactifications on half-flat
  manifolds with $SU(3)$ structure at first order in $\ap$ (ie
  including matter fields). We show that for this model, finding
  solutions with viable gauge group poses the same problems as finding
  solutions in the absence of matter fields.

\vfill

\noindent {December 2008}
  
\end{titlepage}

\section{Introduction}
\label{intro}

Moduli stabilisation in the presence of background fluxes has been an
active field of research in the last years. In particular, several
solutions with stabilised closed string moduli were found. Most
cases include type IIB string compactifications, but examples within
type IIA or heterotic strings as well as M-theory have also been
constructed. By now one can say that there is a fair understanding of
moduli stabilisation in such models. 

On the other hand, this problem has been addressed separately from the
issue of constructing a Standard Model--like sector. Several string
models which reproduce the Standard Model spectrum and gauge group
are known, but moduli stabilisation in these scenarios is still an open
question.  It has been largely assumed (at least in type IIB
scenarios) that the two sectors are independent and one can always add
the Standard Model sector after the closed string moduli have been
stabilised. This procedure was questioned in ref.~\cite{BMP1} where it was
shown that certain constraints have to be fulfilled in order to avoid
the destabilisation of the moduli vacuum by the matter sector. Few
examples have been analysed in the context of type IIB
compactifications in refs.~\cite{BMP2,CMQ}. However, the interaction
between these two sectors still needs a better understanding and
therefore, it is important to address the problem of the moduli
stabilisation in realistic models.

The present note represents a small step in this direction in that we
try to analyse a simple model and see what are the obstacles against a
viable solution with all moduli stabilised in the presence of matter
fields. The model we shall discuss can be obtained from heterotic
string compactifications. The appearance of a visible sector is almost
automatic in such scenarios and therefore the only thing we have to
worry about is actually moduli stabilisation. This is not an easy task
as in heterotic string compactifications the only ordinary flux
available is the NS-NS $H$-flux. Hence, we are forced to take into
account also geometric fluxes ie compactifications on manifolds with
torsion. Here we shall use the models derived in refs.~\cite{GLM1,GLM2,AC}
where so called half-flat manifolds with $SU(3)$ structure and
generalisations thereof were considered. Such models have a $E_6$
gauge group, chiral matter transforming under $\mathbf{27}$ and
$\overline{\mathbf{27}}$ and singlet fields coming from the geometric
moduli of the compactification manifold.  We will show that, a simple
setup which contains only K\"ahler moduli and charged fields
transforming as $\overline{\mathbf{27}}$ under the $E_6$ gauge group,
does not have satisfactory supersymmetric solutions. In particular, we
will see that the problems encountered in finding reliable solutions
can be related to the fact that in the absence of matter fields such
systems do not have satisfactory solutions either.

\section{The model}

As explained in the Introduction, the model we deal with can be
derived from compactifications of the heterotic string on half-flat
manifolds with $SU(3)$ structure \cite{GLM1,GLM2}. We will consider a
simplifying case in that we take $h^{2,1}=0$, where $h^{2,1}$ stands
for the corresponding Hodge number of the original Calabi--Yau
manifold (in other words these manifolds correspond to rigid
Calabi--Yau's and have no complex structure moduli).  The model will
therefore be $N=1$ supergravity coupled to a super Yang-Mills sector
with gauge group $E_6$ with a certain number $h$ of charged fields
$C^i, i =1\ldots h$ transforming as $\overline{\mathbf{27}}$ of $E_6$
and $h$ K\"ahler moduli $T^i$ which are singlets under the gauge
group.\footnote{We ignore the bundle moduli in this analysis.}  In
principle one also has to take into account the dilaton field, but it
will play no role in the present discussion and we will therefore
ignore it for the sake of simplicity. The field $C^i$ carries an
additional index $A$ running over the $27$-dimensional representation
of $E_6$. In the following we shall suppress the index $i$ and
effectively work with $h=1$, but the results can be straightforwardly
generalised to the case $h>1$.

Before we proceed with the model we should note that the case
$h^{2,1}=0$ may not be totally irrelevant from the string
compactification point of view. Suppose we consider a compactification
on a Calabi--Yau manifold with $h^{1,1} > h^{2,1} \ne 0$. This model
will include also complex structure moduli, denoted by $Z$, as well as
charged fields transforming under $\mathbf{27}$ of $E_6$, which we
denote by $D$. In ref.~\cite{GLM2} it was shown that the geometric
fluxes can give mass couplings to the fields $(T,Z)$ and $(C, D)$. By
choosing the flux parameters appropriately we can give large masses to
$h^{2,1}$ pairs of moduli fields $(T, Z)$ as well as to the
corresponding charged fields $(C, D)$, and integrating them out, we
are left precisely with the field content of the model we want to
discuss in this note.

Coming back to the model we want to analyse, the K\"ahler potential
for the chiral fields is given by \cite{GLM2}\footnote{The factors
  which appear in these formulae can be derived from \cite{GLM2}, but
  their precise values are not relevant in the following.}
\begin{equation}
  \label{Kpot}
  K = - 3 \log (T + \bar T) + \ap \frac{6}{(T + \bar T)} C^A \bar C_A +
  \mathcal{O}(\ap^2) \; ,
\end{equation}
and the superpotential
\begin{equation}
  \label{W}
  W = ieT - \frac\ap{3} j_{ABC} C^A C^B C^C + \mathcal{O}(\ap^2)\; .
\end{equation}
The two pieces in this superpotential are not new, but were never
studied simultaneously before. In particular, the cubic term has been
known since the mid eighties, \cite{witten}, but at that
time the moduli superpotential coming from fluxes (including
geometrical ones) was not developed. The linear piece in the
superpotential was studied in ref.~\cite{dCGLM} and it has been concluded
that this piece alone can not solve the stabilisation problem for the
K\"ahler modulus $T$. The main question we want to ask in this note is
whether the addition of the cubic term in the superpotential changes
the conclusions of ref.~\cite{dCGLM}. As we shall see in the following the
problems we encounter in stabilising the moduli in this model can be
traced back to the problems encountered in ref.~\cite{dCGLM}.

\section{Solutions}

The supersymmetric solutions of this system can be found by solving
the F-term equations
\begin{eqnarray}
  \label{FT}
  F_T & = & i e - \frac{3}{T + \bar T} W - \frac{6 \ap C^A \bar C_A}{(T +
    \bar T)^2} W + \mathcal{O}(\ap^2) \; , \\
  \label{FC}
  F_{C^A} & = & \ap \left( - j_{ABC} C^B C^C  + \frac{6 \bar C_A}{(T +
      \bar T)} W \right) + \mathcal{O}(\ap^2) \; .
\end{eqnarray}
The second equation is in general difficult to solve. However, we are not
interested in the most general solution, but we are also looking for
solutions which would be of relevance for the four-dimensional
physics, ie solutions which may include sectors of the standard
model. Therefore it would be interesting to see if there exist
solutions which preserve a gauge group which is large enough to
incorporate the standard model gauge group. Possible solutions within
$E_6$ are: $E_6$ itself, $SO(10) \times U(1)$, $SU(3) \times SU(3) \times
SU(3)$ or $SU(2) \times SU(6)$. 
Under these subgroups $\overline{\mathbf{27}}$ branches as \cite{slansky}
\begin{eqnarray}
  \label{br1}
  \overline{\mathbf{27}} & = & \overline{\mathbf{16}}^1 \oplus \mathbf{10}^{-2}
  \oplus \mathbf{1}^4 \; ; \\
  \label{br2}
  \overline{\mathbf{27}} & = & (\mathbf{3, \bar 3, 1})  \oplus
  (\mathbf{\bar 3, 1, \bar 3}) \oplus (\mathbf{1, 3, 3}) \; ; \\
  \label{br3}
  \overline{\mathbf{27}} & = & (\mathbf{\bar 2, 6}) \oplus (\mathbf{1,
    15}) \; .
\end{eqnarray}
By giving a vev
to the singlet in eq.~\eqref{br1}, $E_6$ is broken to $SO(10)$, which from a
phenomenological point of view, is one of the most interesting GUT
choices. A vev for any of the fields on the RHS of eq.~\eqref{br2} breaks
the gauge group to $SU(3) \times SU(2) \times SU(2)$. Finally, a vev
for the $(\mathbf{\bar 2,6})$ breaks the gauge group to $U(1) \times SU(5)$
while a vev for $(\mathbf{1,15})$ breaks the gauge group to $SU(2)
\times SU(4) \times SU(2)$. We shall analyse these possibilities in
the following subsections.

\subsection{$E_6$ preserving solution}
\label{sec:e_6}

A solution which preserves the $E_6$ gauge group necessarily has
vanishing charged fields, ie $<C^A>=0$. Then equation \eqref{FC} is
automatically satisfied and we are left with eq.~\eqref{FT}. This is a
simplified version of the equations which were analysed in
ref.~\cite{dCGLM} and has no solution for non-vanishing flux parameter
$e$.

Note that there is a possible way around this problem. This however is
not in the main line of this note as it has vanishing matter fields
and, therefore, could have been studied before, but for the sake of
completeness we show why it does not represent a viable solution
either. As we explained in the introduction, the model we consider
here is definitely not a complete one and it can actually be
considered to emerge after integrating out massive fields at a higher
energy scale. In this case one would also have to consider an
additional constant, $w^0$, in the superpotential which represents the
value of the superpotential at the point where the integrated out
fields were stabilised. There is no reason to believe that this
constant is real and therefore equation \eqref{FT} will have the
solution 
\begin{equation}
  \label{Tsol}
  T = - \frac{1}{e} \left( 3 Im \; w^0 - i Re \; w^0 \right) \; .
\end{equation}
Since the flux parameter $e$ is expected to be quantised \cite{dCGLM},
in order to obtain a trustworthy solution for the K\"ahler modulus we
need a large $Im \; w^0$. On the other hand we have to remember that
the full model also contains the dilaton which has to be stabilised as
well. The only known solution is through gaugino condensate in the
hidden sector which can give a large value for the dilaton field when
$W$ is small. Therefore it seems impossible to stabilise
both the K\"ahler moduli and the dilaton at large values in this way.

In conclusion, as expected, $E_6$ preserving solutions are in no
respect different from the solutions found in ref.~\cite{dCGLM}.

\subsection{$SO(10)$ preserving solutions}
\label{sec:so10}

One of the phenomenologically most promising subgroups of $E_6$ is
$SO(10)$. The branching of the $\mathbf{27}$-dimensional
representation of $E_6$ under $SO(10) \times U(1)$ is given in
eq.~\eqref{br1} where the superscripts indicate the $U(1)$
charge. Clearly, giving a vev to the singlet in eq.~\eqref{br1} breaks
the gauge group from $E_6$ down to $SO(10)$. Let us also split the
charged field $C^A$ according to the above branching
\begin{equation}
  \label{Cbr}
  C^A = (C^\alpha , C^a , c) \; ,
\end{equation}
where $\alpha$ and $a$ run over the $\overline{\mathbf{16}}$ and
$\mathbf{10}$ dimensional representations of $SO(10)$ respectively. By
$c$ we have denoted the singlet from eq.~\eqref{br1} and therefore we
are looking for a situation where
\begin{equation}
  \label{Svev}
  <C^{\alpha}> = <C^a> = 0 \; , \qquad <c> \ne 0 \; .
\end{equation}

Since the singlet $c$ is is charged under $U(1)$ a coupling of the
type $c^3$ is forbidden and therefore the only allowed couplings have
the generic form $\overline{\mathbf{16}} \cdot \overline{\mathbf{16}}
\cdot \mathbf{10}$ and $\mathbf{10} \cdot \mathbf{10} \cdot \mathbf{1}$.

Writing eq.~\eqref{FC} for the component field $c$ we obtain
\begin{equation}
  \label{Fc}
  F_c = - \ap j_{ab \mathbf{1}} C^a C^b + K_c W = 0 \; .
\end{equation}
The first term vanishes due to eq.~\eqref{Svev} and since $K_c\sim
\bar c/(T+\bar T)$ can only vanish in the decompactification limit
$T + \bar T \to \infty$, we conclude that $W$ must vanish for such a
solution
\begin{equation}
  \label{W0}
  W=0 \; .
\end{equation}
Clearly, for non-vanishing flux parameter $e$, equation \eqref{FT}
does not have a solution which implies that supersymmetric $SO(10)$
solutions do not exist in this toy model. As anticipated, this problem
can be traced back to the problems encountered in ref.~\cite{dCGLM}. In
particular we see that the absence of supersymmetric
solutions at the zeroth order in $\alpha'$ prevents the existence of
solutions also at the first order in $\alpha'$.

Note that one can think of breaking the gauge group further to
$SU(5)$. This can be done by giving a vev to the singlet in the
branching of $\overline{\mathbf{16}}$ under $SU(5)$. However, since
the couplings $\overline{\mathbf{16}}^3$ or $\overline{\mathbf{16}}^2
\cdot \mathbf{1}$ do not exist either, the above arguments remain
unchanged even for breaking the gauge group to $SU(5)$. Therefore, the
phenomenologically most interesting cases ie $SO(10)$ and $SU(5)$
gauge groups are ruled out in this simple model. It would be
interesting to see if in more complicated models, with more
complicated moduli and matter fields superpotentials, this conclusion
may be changed. It is worth noting that if the condition $W=0$
persists this is still not satisfactory at least from the point of
view of stabilising the dialton for which we need a small, but
non-zero value for the superpotential.

\subsection{$SU(3) \times SU(2) \times SU(2)$ preserving solution}

This solution is intrinsically different from the $SO(10)$
solution. The reason is that the field which breaks $E_6$, say
$(\mathbf{1,3,3})$, admits a coupling of the type
$(\mathbf{1,3,3})^3$. Let's denote this field generically by $B^a$. Then
the corresponding F-term equation will be of the form
\begin{equation}
  \label{FB}
  F_{B^a} = \ap \left( j_{abc} B^b \cdot B^c + \frac{6 \bar B_a}{T +
      \bar T} W \right) \; .
\end{equation}
For a non-vanishing $B$ we necessarily have $W\ne 0$. Note that this
equation is an equation at the first order in $\alpha'$ and therefore
$W$ appearing on the RHS of it is only the flux superpotential term
$eT$. Moreover, the approximation in which the starting theory can be
trusted is one where the charged fields represent small fluctuations
around zero. Therefore in the above equation $B$ has to be taken small
and thus 
\begin{equation}
  \label{Bsol}
  B \sim \frac{6W}{T + \bar T} = \frac{6i e T}{T + \bar T} \ll 1\; . 
\end{equation}
Since the fluxes are quantised this expression can not be made
arbitrarily small. Moreover since the units in which the fluxes are
quantised are of order $1/\sqrt \ap$, this solution can not be trusted
as the $\ap$ expansion we started with in eqs.~\eqref{Kpot} and \eqref{W}
no longer makes sense. At a first glance it may seem that this problem
is not related to the ones encountered in ref.~\cite{dCGLM} when
looking for solutions. However, from \eqref{Bsol} we see that actually
the obstacle in this case is that we can not make $W$ arbitrarily
small while keeping the K\"ahler modulus $T$ large. This was
essentially the major obstruction in ref.~\cite{dCGLM} and was shown
that more complicated moduli superpotentials may alleviate the problem.

\subsection{Other solutions}

The other possible breaking of $E_6$ which we have presented at the
beginning of this section fall in one of the classes discussed above
and we shall only briefly discuss them. In the case of breaking by the
vev of $(\mathbf{\bar 2, 6})$ in eq.~\eqref{br3}, a coupling of the type
$(\mathbf{\bar 2, 6})^3$ does not exist and therefore, like in the
$SO(10)$ case, the F-term equation for the field which acquires a vev
imposes that $W=0$. As we discussed before this can not be a solution
to eq.~\eqref{FT} for non-vanishing flux parameter $e$. The other case
where the breaking of $E_6$ is due to a vev of $(\mathbf{1,15})$ a
cubic coupling for this field exists and therefore one can find a
solution to the system \eqref{FT} and \eqref{FC}, but as noted in the
previous subsection such a solution is not consistent
with all the approximations used in deriving this model.

\section{Conclusions}

In this note we have studied a simple toy model of $N=1$ supergravity
coupled to a visible sector with a gauge group $E_6$ with one chiral
field in the $\overline{\mathbf{27}}$ representation of $E_6$ and a
singlet field. Such models can be obtained from heterotic string
compactifications on manifolds with $SU(3)$ structure (in particular
half-flat manifolds) as described in ref.~\cite{GLM2}. We have analysed
the vacua of this model which have a chance to describe the Standard
Model. The case when the
gauge group $E_6$ is preserved is trivial in the sense that these
solutions were discussed before in ref.~\cite{dCGLM}. For the cases where the
group $E_6$ is broken we have encountered two situations
\begin{enumerate}
\item The field which acquires a vev has a cubic coupling in the
  superpotential. In this case, because the charged fields are
  considered to be small fluctuations around zero, the value for the
  superpotential at the critical point is required to be also small
  which is in tension with the fact that the value for the modulus
  field $T$ has to be large in order to be in a valid regime of the
  supergravity approximation we have been using. We found that this is
  the case for the solutions which break the gauge group to $SU(3) \times
  SU(2)^2$ or $SU(4) \times SU(2)^2$. 

\item The field which acquires a vev does not have a cubic coupling in the
  superpotential. In this case there is simply no supersymmetric
  solution as the F-term equation for the field which gets a vev
  implies a vanishing value for the superpotential which is not
  compatible with the $F_T$ equation \eqref{FT}. We found that
  breaking to $SO(10)$ or $SU(5) \times U(1)$ lie in this class.  
\end{enumerate}

An important feature of the analysis in this note is that the fact
that no satisfactory solutions in the presence of charged fields were
found can be directly related to the fact that, for this model, in the
absence of charged fields, no proper solutions exit, \cite{dCGLM}.
Therefore it is expected that in more general and more realistic
models which also include complex structure moduli and which have
proper solutions at the zeroth order in $\ap$, \cite{dCGLM}, one can
also find trustworthy solutions in the presence of charged fields. In
particular, there will always be $E_6$ preserving solutions which are
merely extensions of the solutions already obtained in
ref.~\cite{dCGLM}. It is also expected that the solutions which fall
in the first class above are better behaved in more complicated models
as it was shown in ref.~\cite{dCGLM} that the moduli can be stabilised
at reasonable values while keeping the superpotential small. It is not
clear what will really happen with the solutions which fall in the
second class above. If $W$ is still required to vanish there will be a
problem in finding solutions for the dilaton from gaugino condensation
in the hidden sector. If however, the conclusion $<W>=0$ can be
avoided this will constitute a completely new case which will have to
be analysed in detail. The analysis of these more complicated models
is left for future work \cite{AM}.

\vspace{1cm}
\noindent
\textbf{Acknowledgments} The author thanks Shanta de Alwis for carefully
reading a previous version of this manuscript. This work was supported
%by the Ministry of Education and Research and 
by the National University Research Council (CNCSIS) under contract
UEFISCSU 3/3.11.2008.

\end{document}